\documentclass[a4paper,aip,jcp,reprint,twocolumn,floatfix]{revtex4-1} %REVTEX4 - style of APS
\usepackage{amsmath,amsthm,latexsym,amssymb,amsfonts,epsfig} 
\usepackage[english]{babel}
\usepackage[utf8]{inputenc}			% utf8 encoding
\usepackage{graphicx,texdraw}   %for figures
\usepackage{color}
\usepackage{overpic} % to write on figures
\usepackage{ulem} % \sout command
\usepackage{newtxtext}
\usepackage{bm}
\usepackage[colorlinks=true, citecolor=blue, linkcolor=blue]{hyperref} % define as last
\newcommand{\lla}{\left\langle}
\newcommand{\rra}{\right\rangle}
\newcommand{\lbra}{\left (}
\newcommand{\rbra}{\right )}

\newcommand{\ehoch}[1]{\exp\left\{{#1} \right\}}

\newcommand{\zjbs}[4]{\textit{#1}, {#4}, \textbf{#2}, {#3}.}

\newcommand{\td}[2]{\frac{\mathrm{d} #1}{\mathrm{d} #2}} %total

\newcommand{\avg}[1]{\left\langle{#1}\right\rangle}

\newcommand{\be}{\mathbf{e}}

\newcommand{\bq}{\mathbf{q}}
\newcommand{\br}{\mathbf{r}}

\newcommand{\BR}{\mathbf{R}}
\newcommand{\BD}{\mathbf{D}}

\newcommand{\BU}{\mathbf{U}}
\newcommand{\BF}{\mathbf{F}}

\newcommand{\BQ}{\mathbf{Q}}

\newcommand{\BH}{\mathbf{H}}
\newcommand{\mi}{\bm{\mu}}

\newcommand{\integr}[3]{\int_{#1}^{#2} \!\!\! \mathrm{d} {#3}\,}

\newcommand{\bee}{\begin{eqnarray}}
\newcommand{\eee}{\end{eqnarray}}

\newcommand{\ez}{\hat{\mathbf{e}}_z}

\begin{document}

\title{Near-Wall Dynamics of Concentrated Hard-Sphere Suspensions: Comparison of Evanescent Wave DLS Experiments, Virial Approximation and Simulations}
\author{Yi Liu}
\affiliation{Forschugszentrum J\"ulich, Institute of Complex Systems ICS-3, J\"ulich, Germany}
\author{Jerzy B{\l}awzdziewicz}
\affiliation{Texas Tech University, Department of Mechanical Engineering, Lubbock, Texas, USA}
\author{Bogdan Cichocki}
\affiliation{Institute of Theoretical Physics, Faculty of Physics, University of Warsaw, Warsaw, Poland}
\author{Jan K. G. Dhont}
\affiliation{Forschugszentrum J\"ulich, Institute of Complex Systems ICS-3, J\"ulich, Germany}
\affiliation{Heinrich-Heine Universit\"at, D\"usseldorf, Germany}
\author{Maciej Lisicki}
\affiliation{Institute of Theoretical Physics, Faculty of Physics, University of Warsaw, Warsaw, Poland}
\author{Eligiusz Wajnryb}
\affiliation{Institute of Fundamental Technological Research, Polish Academy of Sciences, Warsaw, Poland}
\author{Yuan-N. Young}
\affiliation{New Jersey Institute of Technology, Newark, New Jersey, USA}
\author{Peter R. Lang}
\email{p.lang@fz-juelich.de}
\affiliation{Forschugszentrum J\"ulich, Institute of Complex Systems ICS-3, J\"ulich, Germany}
\affiliation{Heinrich-Heine Universit\"at, D\"usseldorf, Germany}

\begin{abstract}
In this article we report on a study of the near-wall dynamics of suspended colloidal hard spheres over a broad range of volume fractions. We present a thorough comparison of experimental data with predictions based on a virial approximation and simulation results. We find that the virial approach describes the experimental data reasonably well up to a volume fraction of $\phi=0.25$ which provides us with a fast and non-costly tool for the analysis and prediction of Evanescent Wave DLS data. Based on this we propose a new method to assess the near-wall self-diffusion at elevated density. Here, we qualitatively confirm earlier results [Michailidou, \textit{et al., Phys. Rev. Lett}, 2009, \textbf{102}, 068302], which indicate that many-particle hydrodynamic interactions are diminished by the presence of the wall at increasing volume fractions as compared to bulk dynamics. Beyond this finding we show that this diminishment is different for the particle motion normal and parallel to the wall.
\end{abstract}

\maketitle

%%%MAIN TEXT%%%%

\section{Introduction}
Soft matter at interfaces is an essential component of many biological, chemical, and industrial processes. The effect of interactions with system boundaries is even more pronounced as modern technology zooms into smaller length scale, where confinement geometry is comparable to particle size. Some interesting examples are antifreeze proteins at water-ice interface\cite{meister2014observation}, 'contact killing' of bacteria by copper surfaces\cite{hans2013role}, design of food with novel texture by structuring water-water interfaces\cite{tromp2014composition}, etc.
The particular case of colloidal particles near a flat solid wall is prominent in the reduced-scale world of micro- and nanofluidics for example in lab-on-chip applications, where colloidal particles may be used to manipulate fluid flow. The phase behaviour and structures formed in colloidal suspensions near walls have been investigated thoroughly using x-ray and neutron scattering techniques with grazing incidence\cite{Lang2004}, while static interactions of colloidal particles with solid surfaces were investigated using e. g. total internal reflection microscopy\cite{Kleschchanok2008}.
On the other hand, experimental investigations of near wall colloidal dynamics have been lagging behind theoretical developments for a long time. The first theoretical considerations of the problem of slow viscous motion of a sphere close to a wall date back to the early twentieth century by Lorentz\cite{Lorentz} and Fax\'en\cite{Faxen}, while first experiments were achieved only in the 1980's\cite{Adamczyk}. Only during the last twenty years, dynamics at interfaces has developed into a major research branch.

The motion of colloidal particles is known to be hindered\cite{HappelBrenner} in the vicinity of a wall due to hydrodynamic interactions (HI). Their effect may be probed by a number of experimental techniques, with the method of choice depending on the system, its size and optical properties. For a review of these methods, see Ref. \cite{Sigel2009}.
Evanescent wave dynamic light scattering (EWDLS) is an important tool to study near-interface dynamics of colloids, and it is the only technique which is available for the investigation of colloids with a size in the 100 nm range. In a typical EWDLS experiment, a laser beam is totally reflected off a glass-solution interface, and an evanescent wave is then created as illumination source. The penetration depth of the evanescent wave can be tuned by varying the incident angle. Particles located in the volume illuminated by the evanescent wave scatter light which is collected by a detecting unit and passed down to a correlator to generate the intensity time autocorrelation function (IACF). Since the method has been devised by Lan {\it et al.}\cite{Lan1986}, it has witnessed rapid development. In early attempts, EWDLS has been applied to study translational diffusion of spherical colloids in dilute solutions\cite{Garnier1991,Ostrowsky1991,Ostrowsky1994,Feitosa1991,Hosoda1998,Lisicki2012}. With a set-up which allows independent variation of the components of the scattering vector $Q_\parallel$, $Q_\perp$, parallel and perpendicular to the surface, respectively, it is possible to distinguish between the anisotropic diffusivity of colloidal spheres in these directions experimentally\cite{Holmqvist2006,Holmqvist2007}. EWDLS has also been employed to study the dynamics of stiff polymers adsorbed to the interface \cite{Loppinet1998}, as well as the collective motion of end-grafted polymer brushes \cite{Fytas1996,Yakubov2004}; near-wall diffusion of a spherical particle in a suspension of rod-like depletants \cite{Holmqvist2007a} and colloidal dumbbells \cite{Haghighi2013}; and rotational diffusion of optically anisotropic spheres \cite{Rogers2012,Lisicki2014}. Notably, evanescent waves have also been used for near-wall nano-velocimetry \cite{Loppinet2012,Li2015}, and to probe dynamics at liquid-liquid interfaces\cite{PhysRevE.83.011601}.

Recent years have brought increasing interest into the effects of confinement on collective dynamics of colloids. To this end, EWDLS experiments have been performed on hard-spheres suspensions with volume fractions up to 42 percent by Michailidou {\it et al.}\cite{Michailidou2009,Michailidou2013}, along with theoretical developments \cite{Swan2011}. On  the basis of a heuristic approximation for the near wall self-diffusivity, these works suggest that for a concentrated suspension, many-particle hydrodynamic interactions are diminished at high volume fractions due to the presence of the wall, which is there referred to as 'screening out'. In this paper we qualitatively confirm this observation. However, we provide a more quantitative method to determine the near-wall self-diffusion coefficients and we are able to show the diminishment of HI affects the self-diffusivity normal and parallel to the interface to a different extent. This becomes possible using the virial approximation for the initial decay rate of the scattered electric field autocorrelation function (EACF), which we described in our earlier contribution \cite{Cichocki2010}. There we presented a detailed discussion of the derivation of exact expressions for the first cumulant (i.e. initial decay rate) of the EACF in a concentrated suspension of hard spheres. After constructing an appropriate theoretical framework based on the Smoluchowski equation, we have presented two methods for practical calculations of the first cumulant: the virial expansion, and precise multipole simulations. While the latter may be used for high accuracy calculations at any volume fraction of the suspension, the virial expansion is expected to correctly reproduce the experimentally measured cumulants up to moderate concentrations.

The aim of this paper is to present our results on near-wall dynamics in a model hard-sphere system, viewed in EWDLS experiments. For the first time we provide a thorough analysis of the scattering vector dependence of the first cumulant, which allows us to assess the range of volume fractions where the virial approximation can be used to describe the experimental data.  By tuning the suspension volume fraction and the penetration depth, we are able to investigate the effect of HI-diminishment for high concentrations and its anisotropy in a more convincing way than proposed earlier\cite{Michailidou2009,Michailidou2013}. Comparing to virial expansion results and simulations, we are able to assess the validity of the former approximate scheme at higher volume fractions. We also discuss in detail the colloidal near-wall self-diffusivity which may be determined from our results in a similar way as proposed for the corresponding bulk property by Pusey\cite{Pusey1978}, Segr\'e\cite{Segre1995} \textit{et al} and Banchio \textit{et al}\cite{Banchio2008} and analysed theoretically by Abade \textit{et al}\cite{Abade2010}.

The paper is organised as follows. After a short review of the theoretical foundations (section \ref{sect:theory}) and the details of the numerical simulations (section \ref{sect:simulation}), we describe the details of sample preparation, the evanescent wave light scattering setup and  the data analysis in section \ref{sect:experimental}. The experimental EWDLS data are compared to the theoretical predictions in the result and discussion section \ref{sect:results} where we also confront our predictions to data published earlier and discuss the progress we make here beyond the state of earlier contributions~\cite{Michailidou2009,Michailidou2013}. Finally we summarize our results in the conclusion section \ref{sect:conclusions}.

\section{Theoretical description}\label{sect:theory}

We consider an ensemble of $N$ identical spherical particles of radius $a$ immersed in a Newtonian solvent of viscosity $\eta$. The fluid is bounded by a planar no-slip wall at $z=0$.

In EWDLS experiments, the scattered light intensity time autocorrelation function $g_2(t)$ is measured, from which the normalised scattered electric field correlation function $\widehat{g}_1(t)$ (EACF) is calculated. Since the scattered electric field $E_s$ depends on the configuration of the system, i.e. the positions of the particles, its fluctuations can be related to the diffusive dynamics of near-wall particles. The initial decay of the EACF is exponential in time
\begin{equation}
\widehat{g}_1(t) = \frac{\avg{E_s(t) E_s^*(t=0)}}{\avg{ |E_s(t)|^2}} \sim \exp(-\Gamma t)\quad\mathrm{as}\ t\to0,
\end{equation}
with $\Gamma$ being the first cumulant, similarly to bulk DLS\cite{BernePecora}. However, there are two important differences to the bulk case. Firstly, the sample is illuminated by a non-uniform evanescent wave. Its intensity decays exponentially with the distance $z$ from the wall as $\exp(-\kappa z)$, thus restricting the scattering volume to a wall-bounded region with a thickness of order $\kappa^{-1}$. The particles staying closer to the boundary receive more intensity and yield the strongest signal. The instantaneous scattered electric field is then given for an ensemble of $N$ particles as\cite{Cichocki2010}
\begin{equation}
E_s \sim \sum_{j=1}^{N} \exp\left(-\frac{\kappa}{2}z_j \right)\exp\left(i\BQ\cdot\br_j\right),
\end{equation}
where $\br_j$ is the position of the centre of sphere $j$, $\BQ$ is the scattering vector and $z_j = \br_j\cdot\ez$, with $\ez$  being a unit vector normal to the wall.

Secondly, the mobility of the particles is strongly hindered by the presence of the wall. The boundary reflects the flow incident upon it, leading to an increase of friction, and thus a slow-down of colloidal dynamics. The effect is more pronounced for particles staying close to the surface where their mobility becomes distance-dependent. This information is encoded in the hydrodynamic mobility tensor $\mi^w_{ij}$ which describes the velocity $\BU_i$ the particle $i$ acquires due to the force $\BF_j$ applied to the particle $j$
\begin{equation}
\BU_i = \mi^w_{ij}\cdot\BF_j.
\end{equation}
For non-interacting spheres in a wall-bounded fluid, the tensors $\mi^w_{ij}$ become diagonal in particle indices, but retain the anisotropic structure which follows from the invariant properties of the system,
\begin{equation}
\mi^w_{ij} = \delta_{ij} [\mu^w_\parallel (\bm{1}-\ez\ez) + \mu^w_\perp\ez\ez].
\end{equation}
where $\bm{1}$ is the unit tensor, and $\mu^w_{\parallel,\perp}$ are scalar mobilities for motion parallel and perpendicular to the boundary. In the absence of the wall, the mobility tensor becomes isotropic, with $\mu_{\parallel}=\mu_\perp=\mu_0 = 1/6\pi\eta a$ being the Stokes mobility of a spherical particle. It follows from the fluctuation-dissipation theorem that the Stokes-Einstein diffusion coefficient $D_0$ is given by $k_B T \mu_0$, where $k_B$ is the Boltzmann constant, and $T$ denotes the temperature. The same relation holds between the many-particle diffusion matrix $\BD$ and the mobility tensor $\mi^{w}$.

Using the Smoluchowski equation formalism, Cichocki {\it et al.}\cite{Cichocki2010} derived an analytical expression for the first cumulant measured in an EWDLS experiment for a suspension of spherical particles,
\begin{equation}\label{Gamma_1}
\Gamma = D_0 \left[\frac{\kappa}{2}\ez - i \BQ\right]\cdot\frac{\BH_w(\kappa,\BQ)}{S_w(\kappa,\BQ)}\cdot\left[\frac{\kappa}{2}\ez + i \BQ\right],
\end{equation}
where the hydrodynamic function reads
\begin{equation}\label{eq:HW}
\BH_w(\kappa,\BQ) = \frac{\kappa}{\mu_0 n A}\sum_{i,j}^N \avg{\exp\left[-\frac{\kappa}{2}(z_i+z_j)\right]\mi^w_{ij}\exp\left[i\BQ\cdot(\br_i - \br_j)\right] },
\end{equation}
and the wall-structure factor is given by
\begin{equation}\label{SW}
S_w(\kappa,\BQ)=\frac{\kappa}{n A} \sum_{i,j}^N \avg{\exp\left[-\frac{\kappa}{2}(z_i+z_j)\right]\exp\left[i\BQ\cdot(\br_i - \br_j)\right] }.
\end{equation}
Here, $n A/\kappa$ is the number of particles within the illuminated scattering volume, with $n$ being the bulk particle number density, and $A$ is the illuminated wall area. The brackets $\avg{\ldots}$ denote ensemble averaging. Eq. (\ref{Gamma_1}) is a generalisation of the bulk result for concentrated suspensions\cite{Gurol}
\begin{equation}
\Gamma = D_0  Q^2 \frac{H(Q)}{S(Q)},
\end{equation}
which corresponds to the limit of infinite penetration depth or $\kappa\to 0$.

Decomposing the scattering vector into components parallel and perpendicular to the wall,
\begin{equation}
\BQ= \BQ_\parallel + \BQ_\perp = Q_\parallel \hat{\be}_\parallel + Q_\perp \ez,
\end{equation}
where $\hat{\be}_\parallel$ is a unit vector in the direction of $\BQ_\parallel$ and using the invariant properties of the system, we arrive at the following structure of the first cumulant
\begin{equation}\label{Gfull}
\Gamma= \frac{D_0}{S_w} \left[\left(\frac{\kappa^2}{4}+ Q_\perp^2\right) H_\perp + Q_\parallel^2 H_\parallel + \frac{\kappa}{2}Q_\parallel  H_I + Q_\parallel Q_\perp H_R \right],
\end{equation}
where
\begin{align}\label{HV}
H_\perp &= \ez \cdot \BH_w(\kappa,\BQ)\cdot \ez, \\
H_\parallel  &= \hat{\be}_\parallel \cdot \BH_w(\kappa,\BQ) \cdot \hat{\be}_\parallel, \\
H_I &= \ez \cdot  2\mathrm{Im}[\BH_w(\kappa,\BQ)]\cdot \hat{\be}_\parallel, \\ \label{HR}
H_R &= \hat{\be}_\parallel \cdot  2\mathrm{Re}[\BH_w(\kappa,\BQ)]\cdot \ez,
\end{align}
with $\mathrm{Im}$ and $\mathrm{Re}$ standing for the imaginary and real part, respectively. The coefficients $H$ as well as $S_w$ may be either evaluated numerically using the virial expansion approach, or by extracted from numerical simulations. Both techniques are briefly described in the course of this work. The expressions given above are valid for an arbitrary wall-particle interaction potential. Further on, we restrict to hard-core interactions.

In the dilute regime, the hydrodynamic function and structure factor have only single-particle contributions, from which it follows that $H_I = H_R = 0$. The surviving parts $D_0 H_\parallel/S_w$ and $D_0 H_\perp/S_w$ in Eq. (\ref{Gfull}) simplify then to the single-particle average diffusion coefficients $\avg{D_\parallel}_\kappa$ and $\avg{D_\perp}_\kappa$, respectively, in agreement with the notation proposed in earlier works\cite{Holmqvist2006,Holmqvist2007,Lisicki2012}. In the case of hard-core sphere-wall interactions, the penetration-depth average (in the dilute limit) reads
\begin{equation}\label{davg}
\avg{\ldots}_\kappa = \kappa \integr{a}{\infty}{z} e^{-\kappa (z-a)} (\ldots).
\end{equation}
We may now explicitly write the first cumulant in this case as\cite{Holmqvist2006,Holmqvist2007}
\begin{equation}\label{G_dilute}
\Gamma = Q_\parallel^2 \avg{D_\parallel}_\kappa + \left(\frac{\kappa^2}{4} + Q_\perp^2\right)\avg{D_\perp}_\kappa.
\end{equation}
The averaged diffusivities $\avg{D_{\parallel,\perp}}_\kappa$ have been calculated as functions of $\kappa a$ in Ref. \cite{Lisicki2012}.

Importantly, this is also the case in the limit of $Q_\parallel\to\infty$ or $Q_\perp\to\infty$, where only the self-parts of $\BH_w$ and $S_w$ survive. The cumulant may then be expressed using the self-diffusion tensor $\BD^s$ which is defined as the initial slope of the mean square displacement tensor of a tracer particle located at a height $z$ at $t=0$, viz.
\begin{equation}
\BD^s(z) = \frac{1}{2}\td{}{t} \avg{
\Delta\br(t) \Delta\br(t)}_{t=0},
\end{equation}
where $\Delta\br(t)$ is the displacement vector of the tracer particle during the time $t$. The tensor $\BD^s$ may be expressed in terms of the mobility matrix $\mi^w$, as we have shown in Ref. \cite{Cichocki2010}. Thus, the cumulant may be approximated for sufficiently large $Q_\parallel$ or $Q_\perp$ by
\begin{equation}\label{Gamma_self}
\Gamma \approx Q_\parallel^2 \avg{D^s_\parallel}_\kappa + \left(\frac{\kappa^2}{4} + Q_\perp^2\right)\avg{D^s_\perp}_\kappa,
\end{equation}
where $D^s_{\parallel,\perp}$ are the components of the self-diffusion tensor $\BD^s(z)$, with the average given by
\begin{equation}\label{selfavg}
\avg{D^s_{\parallel,\perp}}_\kappa = \frac{\displaystyle \integr{0}{\infty}{z} e^{-\kappa z} g(z) D^s_{\parallel,\perp}(z)}{\displaystyle \integr{0}{\infty}{z}  e^{-\kappa z} g(z)},
\end{equation}
and $g(z)$ being the single-particle distribution function. Its definition reads
\begin{equation}\label{eq:ngofz}
n g(z)= N \integr{}{}{\br_2}\ldots\integr{}{}{\br_N} P^w_\mathrm{eq}(\BR),
\end{equation}
where $P^w_\mathrm{eq}(\BR)$ is the equilibrium probability density function (in the presence of a wall) for the system to be at a configuration $\BR=\{\br_1,\ldots,\br_N\}$. The quantities in Eqs. (\ref{eq:HW}), (\ref{SW}), and (\ref{eq:ngofz}), are taken in the thermodynamic limit, which has been discussed for a wall-bounded system in Ref. \cite{Cichocki2010}. In a dilute system with hard sphere-wall interactions, and when interactions between the particles may be neglected, the average (\ref{selfavg}) reduces to the formula (\ref{davg}).

\section{Virial expansion}
For moderately concentrated systems, calculations of the wall
structure factor $S_{w}$ and the components of the wall-hydrodynamic tensor $\mathbf{H}_{w}$ may be performed by expanding them in terms of powers of bulk-particle
concentration $n$ far from the wall. The procedure has already been elaborated in great detail in Ref. \cite{Cichocki2010}. Thus, we refrain here from the technical aspects, focusing on the resulting expressions.

The small dimensionless parameter in the density expansion is the bulk volume
fraction,
\begin{equation}
\phi =\frac{4\pi }{3}a^{3}n,  \label{060a}
\end{equation}%
instead of the concentration $n$. The virial expansion of the wall-structure factor (\ref{SW}) reads
\begin{equation}\label{S_vir}
S_{w}(\kappa ,\mathbf{q})=S^{(1)}(\kappa )+\phi S^{(2)}(\kappa ,\mathbf{q})+\mathcal{O}(\phi ^{2}).
\end{equation}
The coefficient $S^{(1)}$ and the self-part of $S^{(2)}$ may be found analytically as
as
\begin{align}\label{S1self}
S^{(1)} &= e^{-\kappa a}, \\ \label{S2self}
S^{(2)}_\mathrm{self} &=  \frac{2 e^{-\kappa a}}{(\kappa d)^3}[6-3(\kappa d)^2+2(\kappa d)^3-6 e^{-\kappa d}(1+\kappa d)].
\end{align}
with the particle diameter $d=2a$. The distinct part of $S^{(2)}$ has to be evaluated numerically. The analogous virial expansion of the wall hydrodynamic tensor requires a cluster decomposition of the mobility matrix\cite{Cichocki1999}, and has a similar form
\begin{equation}\label{H_vir}
\mathbf{H}_{w}(\kappa ,\mathbf{Q})=\mathbf{H}^{(1)}(\kappa)+\phi
\mathbf{H}^{(2)}(\kappa ,\mathbf{Q})+\mathcal{O}(\phi ^{2}).
\end{equation}
In this case in order to obtain the terms $\mathbf{H}^{(1)}$ and $\mathbf{H}^{(2)}$ we need the
one- and two-particle cluster components of the mobility matrix. Explicit expressions for $S^{(1),(2)}$ and $\BH^{(1),(2)}$ are rather complex, and have been given explicitly in Ref. \cite{Cichocki2010}. To calculate
them, the {\sc Hydromultipole} code, implemented according to Ref. \cite{Cichocki2000}, has been used.

Inserting the  expansions (\ref{S_vir}) and (\ref{H_vir}) into Eq. (\ref{Gamma_1}), we find the following virial expansion for the first cumulant
\begin{equation}
\Gamma = \Gamma^{(1)}(\kappa,\BQ) + \phi \Gamma^{(2)}(\kappa,\BQ) + \mathcal{O}(\phi ^{2}),
\end{equation}
where the factor
\begin{equation}
\Gamma^{(1)}(\kappa,\BQ) = D_0 \left[\frac{\kappa}{2}\ez - i \BQ\right]\cdot\frac{\BH^{(1)}}{S^{(1)}}\cdot\left[\frac{\kappa}{2}\ez + i \BQ\right],
\end{equation}
is the infinite dilution prediction, given explicitly by Eq. (\ref{G_dilute}), while the second term reads
\begin{equation}
\Gamma^{(2)}(\kappa,\BQ) = D_0 \left[\frac{\kappa}{2}\ez - i \BQ\right]\cdot\frac{ \BH^{(2) }S^{(1)}-\BH^{(1)}S^{(2)}}{(S^{(1)})^2}\cdot\left[\frac{\kappa}{2}\ez + i \BQ\right].
\end{equation}
These virial expansion results, together with simulations that are also valid at high concentrations, will be compared to experiments in section \ref{sect:results}. The relations above may be transformed using Eqs. (\ref{HV})--(\ref{HR}) and expressed in terms of the tensorial components of the hydrodynamic function $\BH_w$. The subsequent section contains the details of simulations.

\section{Numerical simulations}\label{sect:simulation}

To determine the equilibrium wall-structure factor (\ref{SW}), the
hydrodynamic functions (\ref{HV})--(\ref{HR}), and the first cumulant
(\ref{Gamma_1}), we have carried out a series of numerical simulations
for a wall-bounded hard-sphere system with particle volume fractions
in the range $0<\phi\le0.3$.  Key elements of our numerical techniques
are summarized below; a more detailed description is provided in our
previous paper\cite{Cichocki2010}.

Since hydrodynamic-interaction algorithms are unavailable for a
single-wall system with periodic-boundary conditions, the calculations
were performed for a suspension confined between two well separated parallel walls.
The equilibrium particle distributions were determined using a
standard Monte--Carlo (MC) algorithm, and the multiparticle mobility matrix
$\boldsymbol{\mu}_{ij}^w$ was evaluated using the periodic version
\cite{Blawzdziewicz-Wajnryb:2008} of the Cartesian-representation
algorithm
\cite{Bhattacharya-Blawzdziewicz-Wajnryb:2005,
Bhattacharya-Blawzdziewicz-Wajnryb:2005a,
Bhattacharya-Blawzdziewicz-Wajnryb:2006%
}
for a suspension of spheres in a parallel-wall channel.

Most of our calculations were carried out for a wall separation $h=13d$
(where $d$ is the sphere diameter).  By
comparing results for different values of $h$, we have established
that the above wall separation is sufficient to obtain accurate one-wall
results, provided that the particle volume fraction is adjusted for
the excess particle density in the near-wall regions.

To evaluate the required volume-fraction correction, we constructed
the equilibrium ensemble for a reference system with a large wall
separation $h=h_0$ and the assumed particle number density $n$ in
the middle of the channel.  The excess particle number per unit area, $n_\mathrm{ex}$, was determined using the formula
\begin{equation}
\label{excess_equation}
N=Ah n+2A n_\mathrm{ex},
\end{equation}
where $N$ is the number of particles in the periodic cell, $A$ is the
wall area, and $h=h_0$ is the wall separation in the reference system.
The particle number $N=N(h)$ for channels with different widths $h$ is
obtained from expression (\ref{excess_equation}), with known
reference values of $n$ and $n_\mathrm{ex}$.

Since the evanescent wave scattering occurs only near the illuminated
surface, and the hydrodynamic field associated with the periodic
forcing $\sim\exp(i \BQ\cdot\mathbf{r})$
decays on the length scale $l\sim Q_\parallel^{-1}$ with the distance
from the wall, the effect of the second wall of the channel on the
multiparticle mobility is small\cite{Bhattacharya-Blawzdziewicz-Wajnryb:2005a}.  We find that for the evanescent wave
parameters corresponding to our experiments, the effect of the second
wall on the hydrodynamic functions (\ref{HV})--(\ref{HR}) is
smaller than the statistical simulation inaccuracies.

The hydrodynamic tensor $\BH_w$ was determined as an average over $M$
independent MC trials.  To obtain statistical accuracy of the order of 2\,\%,
we have used $M$ in the range from $M=30$ for large systems with
$N\approx10^3$ particles to $M=400$ for $N\approx 200$ particles.

\section{Experimental details}\label{sect:experimental}
\subsection{Hard-sphere sample and preparation}
As model systems for the EWDLS experiments, we used two batches of poly (methyl methacrylate) (PMMA) particles, named ASM470 and ASM540 in the following, which were purchased from Andrew Schofield, University of
Edinburgh. The spherical particles are covered with a thin poly-12-hydrostearic acid layer to stabilize them against aggregation in organic solvents. To allow scattering experiments at high volume fractions the particles were transferred from a cis-decaline suspension (as received) to a refractive index matching cis-decaline/tetraline mixture by spinning and re-dispersing them. The solvent used had a cis-decaline mass fraction of $w=0.2$, a refractive index of $n_2 = 1.498$ and a viscosity of $\eta=2.658$ mPas at temperature of $T=298$ K as measured using an Abbemat RXA156 and an Automated Microviscometer AMV$\eta$ from Anton Paar, Graz, Austria.
To determine the particle radius, we employed standard Dynamic Light Scattering (DLS) measurements. The recorded time autocorrelation functions of the scattered intensity $g_2(t)$ (IACF) were analysed by three different methods, namely cumulant analysis, stretched exponential fitting and inverse Laplace transformation. The three methods yield hydrodynamic radii of $R_H = 98$ nm (ASM470) and $R_H = 144$ nm (ASM540) with a variation of less than 1 nm in both cases. These values are assumed to be identical with the hard sphere particle radius $a$ in the following.  Further, the size distributions obtained from inverse Laplace transformation showed a full width at half maximum of less than five percent. The negligible size polydispersity is confirmed by the observation that the suspensions crystallize at sufficiently large particle volume fractions.

Prior to the scattering experiments, the suspensions were filtered through PTFE syringe filters with a nominal pore size of 1 $\mu$m directly into the measurement cells to minimize parasitic scattering from dust particles. To reduce the number of necessary alignment processes of the EWDLS measuring cell, this was filled with the hard sphere suspension of highest volume fraction, and further dilution was achieved by removing a part of the sample and replacing it by pure solvent. The exact volume fraction was determined a posteriori by drying a 250 $\mu$l aliquot and determining the mass of the remaining particles. Further, the EWDLS sample cell was equipped with a small magnetic stirrer bar with which the samples were homogenised before each angular scan to minimize the influence of particle sedimentation.

\subsection{EWDLS set-up}\label{section:setup}
\begin{figure}[ht]
  \begin{center}
   \includegraphics[width=0.5\columnwidth]{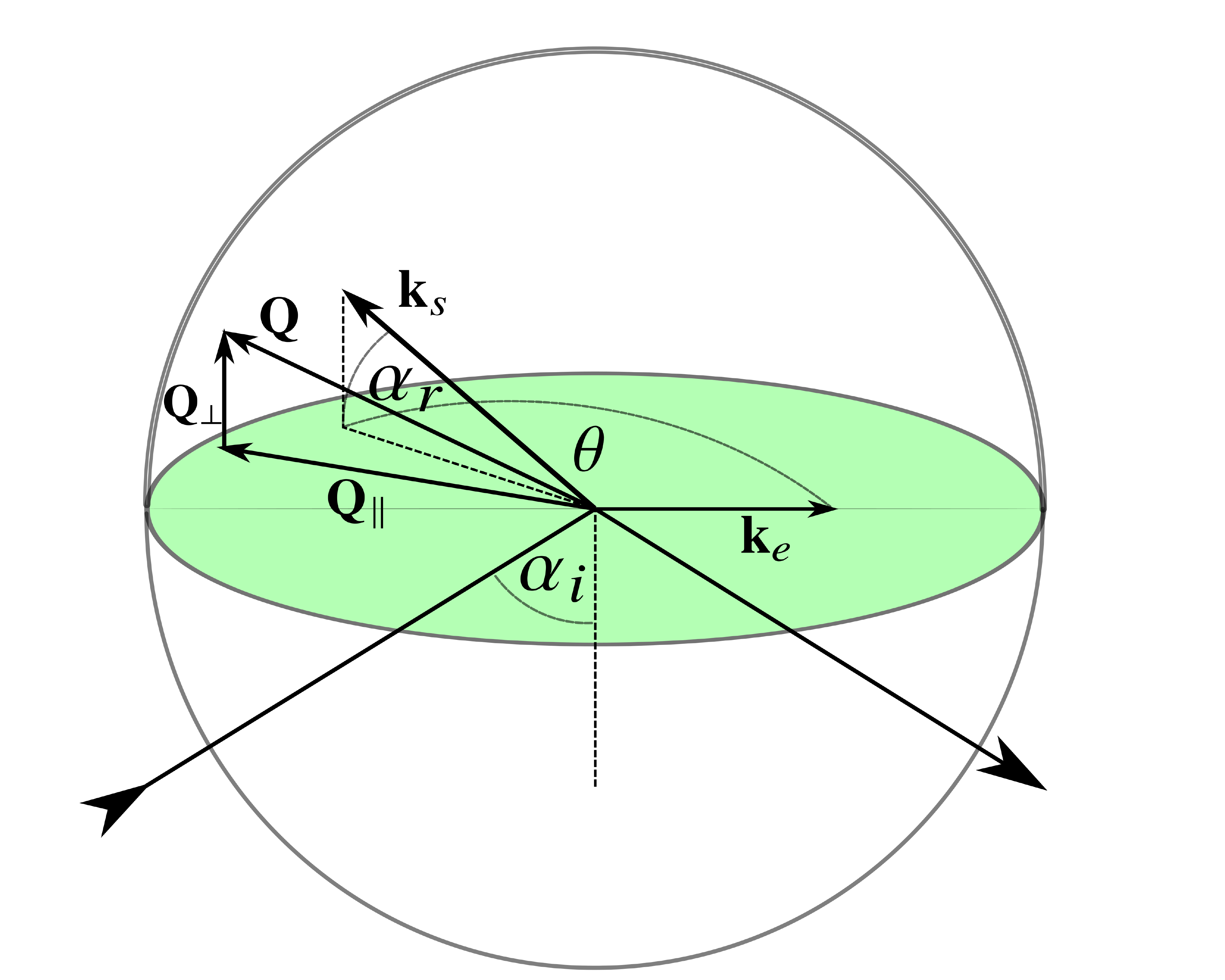}\includegraphics[width=0.5\columnwidth]{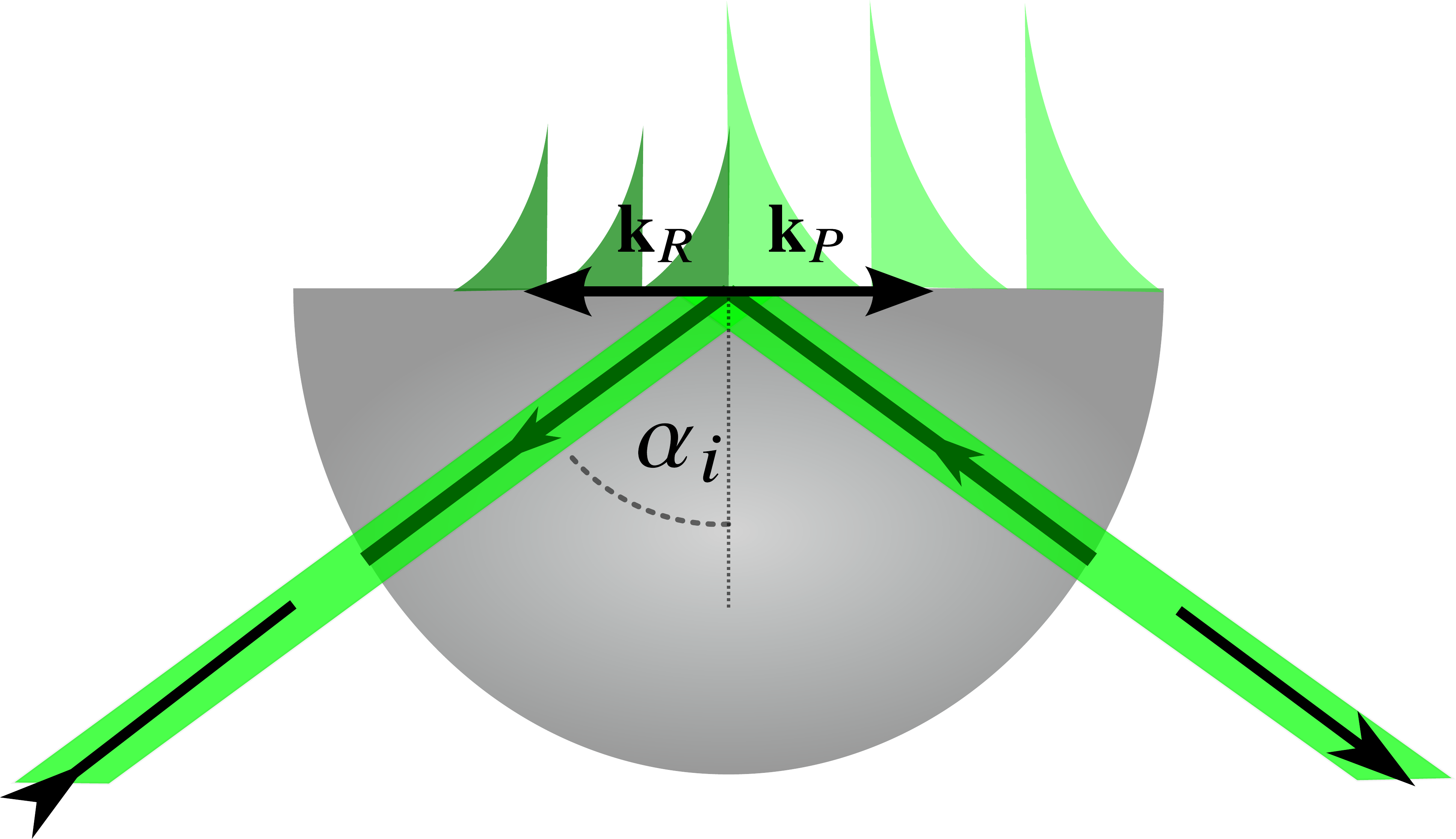}
   \caption{Scattering geometry in EWDLS setup with a spherical geometry. Left: For the definition of angles and wave vectors. Right: For the illustration of the primary beam being back reflected at exit of the hemispherical lens to air, thereby creating a second evanescent wave.}\label{scattering_geometry}
  \end{center}
\end{figure}

EWDLS experiments were performed with a home-built instrument, based on a triple axis diffractometer by Huber Diffraktionstechnik, Rimsting, Germany, which has been described in detail elsewhere\cite{Holmqvist2007}. The setup is equipped with a frequency doubled Nd/Yag Laser (Excelsior; Spectra Physics) with a vacuum wavelength of $\lambda_0 = 532$ nm and a nominal power output of 300 mW as a light source. Scattered light is collected with an optical enhancer system by ALV Lasververtriebsgesellschaft, Langen, Germany, which is connected to two avalanche photo diodes by Perkin Elmer via an ALV fiber splitter. The TTL signals of the diode were cross-correlated using an ALV-6000 multiple tau correlator.
The scattering geometry and the definition of the scattering vector and its component parallel and normal to the interface are sketched in Fig. \ref{scattering_geometry}.
The sample cell (custom-made by Hellma GmbH, Muellheim, Germany) consists of a hemispherical lens as the bottom part, made of SF10 glass, with an index of refraction
$n_1 = 1.736$ at $\lambda_0 = 532$ nm. The hard sphere suspension is contained in a hemispherical dome sitting on top of the lens. The primary beam is reflected off the interface between the
glass and the solution, by that creating an evanescent wave in the the solution which is used as the illumination for the scattering experiment. The evanescent wave has a wave vector $\mathbf{k}_e$ pointing in the direction of the reflected beam's projection onto the reflecting interface. The scattering vector is given by $\mathbf{Q}=\mathbf{k}_s-\mathbf{k}_e$, where the scattered light wave vector, $\mathbf{k}_s$, is defined by the two angles $\theta$ and $\alpha_r$ which describe the position of the detecting unit.

The inverse penetration depth of the evanescent wave depends on the angle of incidence  $\alpha_i$ as $\kappa/2=2\pi\sqrt{(n_1\sin \alpha_i)^2-n_2^2}/\lambda_0$. The magnitudes of the scattering vector components parallel $Q_\parallel=2\pi n_2 \sqrt{1+\cos^2\alpha_r-2\cos\theta\cos\alpha_r}/\lambda_0$ and normal $Q_\perp=2\pi n_2\sin\alpha_r/\lambda_0$ to the interface can be varied by changing $\theta$ and $\alpha_r$. In a typical EWDLS experiment, we record series of correlation functions at fixed $Q_\perp$ varying $Q_\parallel$ ($Q_\parallel$-scan) or vice versa ($Q_\perp$-scan).

\subsection{Details of data analysis}\label{sect:data_analysis}

The analysis of the scattered intensity time autocorrelation function $g_2(t)$ from EWDLS is much less straightforward than in conventional bulk dynamic light scattering (DLS), mainly for two reasons. The first major complication occurs from the fact that a simple quadratic Siegert relation between $g_2(t)$ and the correlation function of the scattered field $g_1(t)$ which is usually assumed in DLS does not apply in most cases in EWDLS. As described in section \ref{section:setup}, the incident laser beam is totally reflected from the glass/solution interface in EWDLS. As this interface is always corrugated, it acts as a static scatterer which in general contributes significantly to the observed signal. Therefore a mixed homodyne/heterodyne detection scheme has to be taken into account, and the generalized Siegert relation\cite{dhon96}
\begin{equation}\label{g2g1}
g_2(t)=1+2C_1\widehat{g}_1(t)+(C_2\widehat{g}_1(t))^2
\end{equation}
for the conversion from $g_2(t)$ to $\widehat{g}_1(t)$ has to be
applied. Here, $C_2=1-\sqrt{1-A}$ and $C_1=C_2-C_2^2$, with $A$ being
the experimental intercept of $g_2(t)$.
Further, in many cases EWDLS intensity-autocorrelation functions exhibit a
very slow decay at large times. The physical origin of this slow
relaxation is not clear yet. While Garnier {\it et al.} conjecture that it is
due to a slow reversible adsorption of the particles to the wall
due to van der Waals attraction~\cite{Garnier1991}, Steffen \cite{stef12} and Lisicki {\it et al.}\cite{Lisicki2012} argue that it is also caused by the unavoidable
stray-light from surface defects, which is scattered
by colloids in the bulk of the suspension into the detector. Since these slow
modes are in general well separated from the relaxation rates of interest, we approximate their contribution by an
additional baseline $B_1$ to $\widehat{g}_1(t)$.

Thus, to determine the initial slope $\Gamma$ of $\widehat{g}_1(t)$, which is related to the dynamic properties of interest, we chose to
non--linear least squares fit  the experimental correlation
functions to Eq.~(\ref{g2g1}), where $g_1(t)$ is modelled as a decaying single exponential function in time
\begin{equation}\label{g2g1plusbl}
\widehat{g}_1(t)=(1-B_1)\ehoch{-\Gamma t}+B_1.
\end{equation}
According to Eqs.~(\ref{g2g1}) and (\ref{g2g1plusbl}), $B_1$ is
related to $B_2$, the baseline of $g_2(t)$,  by
$B_1=\sqrt{(C_1/C_2^2)^2+B_2/C_2^2}-C_1/C_2^2$. Consequently there
are three fit parameters $A$, $\Gamma$ and $B_2$. Since an
erroneous baseline value will lead to a deviation of $\Gamma$ from
its true value, due to a normalization error, we fitted
$g_2(t)$ repeatedly, starting with a number of data points, $N_p$. After a single fit had converged, two data points at the long time end of $g_2(t)$ were removed, reducing $N_p$ by two, and the remaining data points were fitted again. This procedure was repeated until $N_p<20$. With this technique it was possible to identify a limited range of $N_p$'s where the values of $B_2$ and $\Gamma$ are essentially independent of $N_p$.
The $\Gamma$ values determined in this range are considered to be the initial slope or the first cumulant of $\widehat{g}_1(t)$. Where error bars are presented with values of $\Gamma$, they reflect the standard deviation of repeated measurements.

\subsection{Effect of back-reflection}\label{sect:backrefl}
An additional difficulty in EWDLS stems from the fact that, different from bulk DLS, it is not possible to apply a refractive index-matching batch around the sample cell. Therefore the primary beam will be back-reflected at the exit from the semi-spherical lens with a reflectance $R$. In the present case the semi-spherical lens has a refractive index of $n_1= 1.736$, which leads to a reflectance of $R=0.072$ according to Fresnel's equations\cite{BornWolf}. As sketched in Fig.~\ref{scattering_geometry}, the back reflected beam will also be reflected off the glass sample interface, thereby causing a second evanescent wave with wave vector $\mathbf{k}_R=-\mathbf{k}_P$, where $\mathbf{k}_P$ is the wave vector of the evanescent wave caused by the original primary beam. In what follows, the subscript  $_P$ will refer to the evanescent wave caused by the primary beam, while $_R$ will be associated with the evanescent wave due to the back reflected beam. The latter gives rise to  a second scattering process, for which the in-plane scattering angle is $\theta_R=180-\theta_P$. Consequently the scattering vector components parallel to the interface are given by
\begin{equation}
Q_{\parallel,i} =\frac{2\pi}{\lambda_0}n_2\sqrt{1+\cos^2{\alpha_r}-2\cos{\alpha_r}\cos{\theta_i}},
\end{equation}
where $i\in \{R,P\}$. Differently, the component normal to the interface remains unchanged in the two cases
\begin{equation}
Q_{\perp,R}=Q_{\perp,P}\doteq Q_\perp=\frac{2\pi}{\lambda_0}n_2\sin\alpha_r.
\end{equation}
The normalized field correlation functions in such a situation should be considered as a weighted sum of two individual correlation functions from two scattering experiments
\begin{equation}\label{backrefg1}
\widehat{g}_1(t) = \frac{P(Q_P)}{P(Q_P)+RP(Q_R)}g_1^P(t)+R\frac{P(Q_R)}{P(Q_P)+RP(Q_R)}g_1^R(t),
\end{equation}
where $Q_P=\sqrt{Q^2_{\parallel,P}+Q^2_\perp}$, $Q_R=\sqrt{Q^2_{\parallel,R}+Q^2_\perp}$ and $P(Q_i)$ is the particle scattering factor of a sphere.

In evanescent illumination, the scattering factor is affected by the non-uniform character of the electric field and becomes penetration-depth dependent. For an optically uniform particle, the scattering amplitude in the evanescent field reads $B(\BQ,\kappa) = \frac{1}{V}\int_{V} \exp\left[(i\bq+\tfrac{\kappa}{2}\ez)\cdot\br\right] \mathrm{d}\br$.  Thanks to the high symmetry, for a spherical particle of radius $a$, $B(\BQ,\kappa)$ can be explicitly calculated as
\begin{equation}
B(\BQ,\kappa) = 3 \left[\frac{ca  \cosh(ca) - \sinh(ca)}{(ca)^3}\right],
\end{equation}
with $c = \sqrt{ -Q^2 - i Q_\perp \kappa + \frac{\kappa^2}{4}}$. The particle scattering factor is then found as $P(Q)=|B(\BQ,\kappa)|^2$.

In order to illustrate the effect of back-reflection for a dilute suspension, we analyse Eq. (\ref{backrefg1}) using the field correlation function given by the first two cumulants:
\begin{equation}
g_1^i(t) \approx \exp\left(-\Gamma_{i}t+\frac{1}{2}\Gamma_{2,i}t^2\right),
\end{equation}
with the first cumulant $\Gamma_{i}$ given in the dilute regime by Eq. (\ref{G_dilute}), and the second cumulant can be calculated as
\begin{equation}
\Gamma_{2,i}=\gamma_{i}-\Gamma_{i}^2, \label{gamma2}
\end{equation}
where the second moment $\gamma_i$ is defined as\cite{Lisicki2012}
\begin{align}\label{mu2}
\gamma_{i}&=Q_{\parallel,i}^4\lla D^2_\parallel\rra_\kappa+\lbra Q^4_\perp-\frac{\kappa^4}{16}\rbra \lla D^2_\perp\rra_\kappa\\\nonumber
&+2Q_{\parallel,i}^2\lbra Q^2_\perp-\frac{\kappa^2}{4}\rbra\lla D_\parallel D_\perp\rra_\kappa\\\nonumber
&+\kappa Q_{\parallel,i}^2\lla D'_\perp D_\parallel\rra_\kappa+\lbra Q^2_\perp+\frac{\kappa^2}{4}\rbra \lla \lbra D'_\perp\rbra^2\rra_\kappa.
\end{align}
Here $D'_\perp = \td{ }{z}[D_\perp (z)]$. The resulting IACF has to be calculated from $\widehat{g}_1(t)$ using the generalized Siegert relation, Eq. (\ref{g2g1}). The averaged diffusion coefficients, which are required for the calculations of $\Gamma$ and $\Gamma_2$ at a given value of $\kappa a$ were calculated in Reference \cite{Lisicki2012}.

\begin{figure}[ht]
  \begin{center}
     \includegraphics[width=0.8\columnwidth]{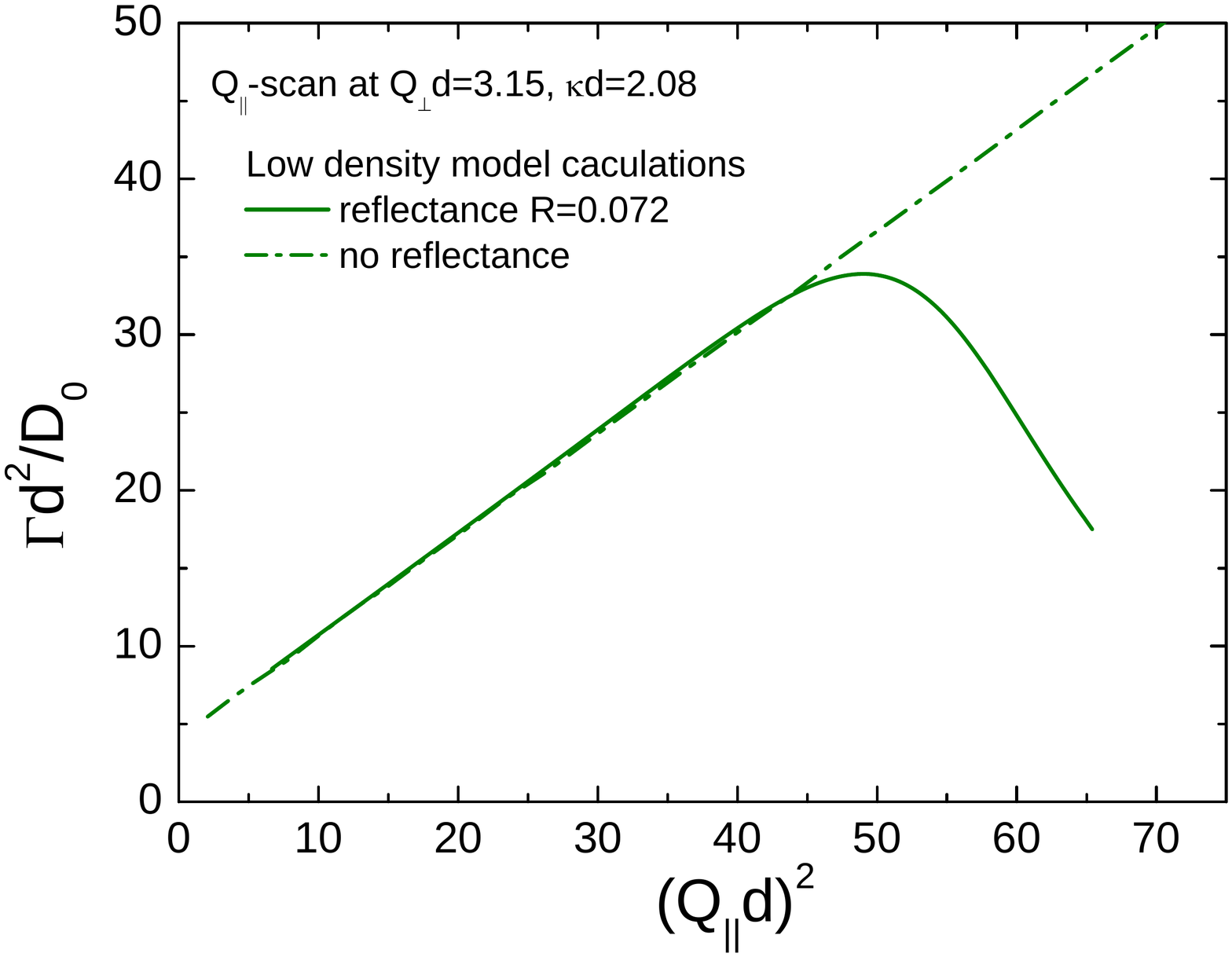}
   \includegraphics[width=0.8\columnwidth]{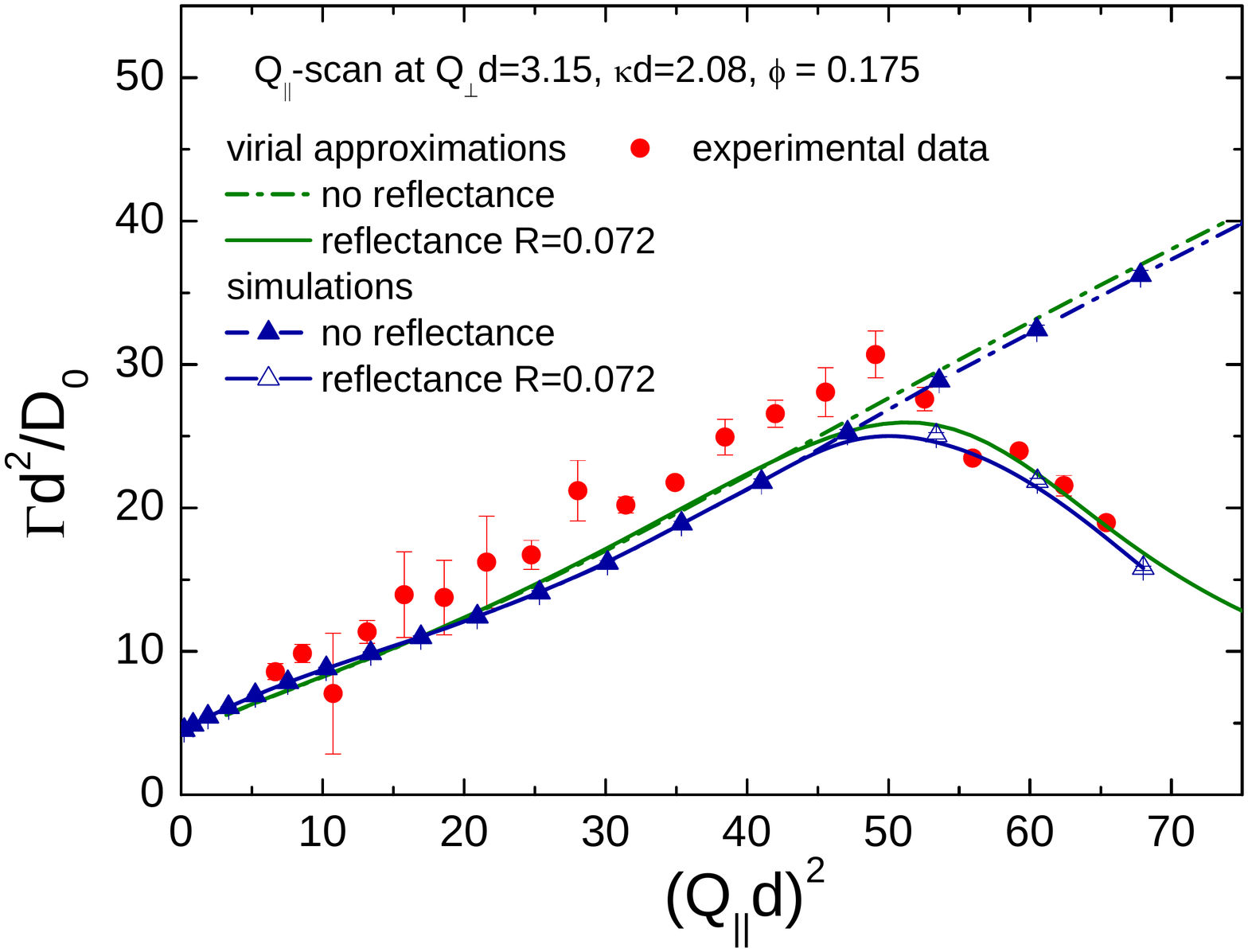}
   \caption{Top: Result of model calculations for zero particle density  without (dashed dotted line) and with (full line) taking into account the effect of back-reflection, showing a considerable difference in the high-$Q$ range. Bottom: Comparison of experimental data obtained at $\phi=0.175$ (full circles) to simulations (line with triangles) and and virial approximations at the same concentrations with (full line) and without (dashed dotted line) taking into account the effect of back reflection. The experimental parameter $Q_\perp d = 3.15$ and $\kappa d= 2.08$ are the same for both graphs.}\label{fig:model_calc_br}
  \end{center}
\end{figure}

The calculated model correlation functions are now evaluated according to the same analysis procedure as the experimental data to obtain initial slopes $\Gamma$ as a function of $Q_\parallel$. In the top part of Fig. \ref{fig:model_calc_br}  we compare initial relaxation rates of model correlation function, which were calculated in this way at infinite dilution. The model calculations coincide perfectly at low scattering vectors. However, those data, which were calculated taking into account the effect of back-reflection, strongly decrease at larger scattering vectors. Here and in the following, we will present the results in dimensionless form, i. e. relaxation rates in units of $D_0/d^2$ and scattering vectors in units of $1/d$ where $D_0$ is the particles' bulk diffusion coefficient at infinite dilution.

For a concentrated suspension, both $\Gamma_P$ and $\Gamma_R$ may be calculated from the virial expansion and from simulations. The first cumulant of the EACF including the back-reflection effect may thus be written from Eq. (\ref{backrefg1}) as
\begin{equation}\label{eq:gamma_with_br}
\Gamma = \frac{P(Q_P)}{P(Q_P)+RP(Q_R)}\Gamma_P+R\frac{P(Q_R)}{P(Q_P)+RP(Q_R)}\Gamma_R,
\end{equation}
again without any free parameter. In the bottom part of Fig.~\ref{fig:model_calc_br} the same experimental data are compared to simulation results and to virial calculations for $\phi=0.175$. In both, the virial calculations and the simulations, the effect of back-reflection can be included as described above. It turns out that up to $Q_\parallel d\sim 7$ the first term in Eq. (\ref{eq:gamma_with_br}) dominates, so that $\Gamma\approx \Gamma_P$, and the back-reflection effect need not be taken into account. However, in the high-$Q$ range, the back-reflection is essential to correctly reproduce the experimental data, as seen in Fig.~\ref{fig:model_calc_br}.
We are therefore led to conclude that the first cumulants obtained experimentally at high in-plane angles, i. e. $\theta >\pi/2$ should be considered with extreme care and potentially discarded when comparing experimental data to theoretical predictions and simulations.

\section{Results and Discussion}\label{sect:results}
\begin{figure}[ht]
  \begin{center}
   \includegraphics[width=0.8\columnwidth]{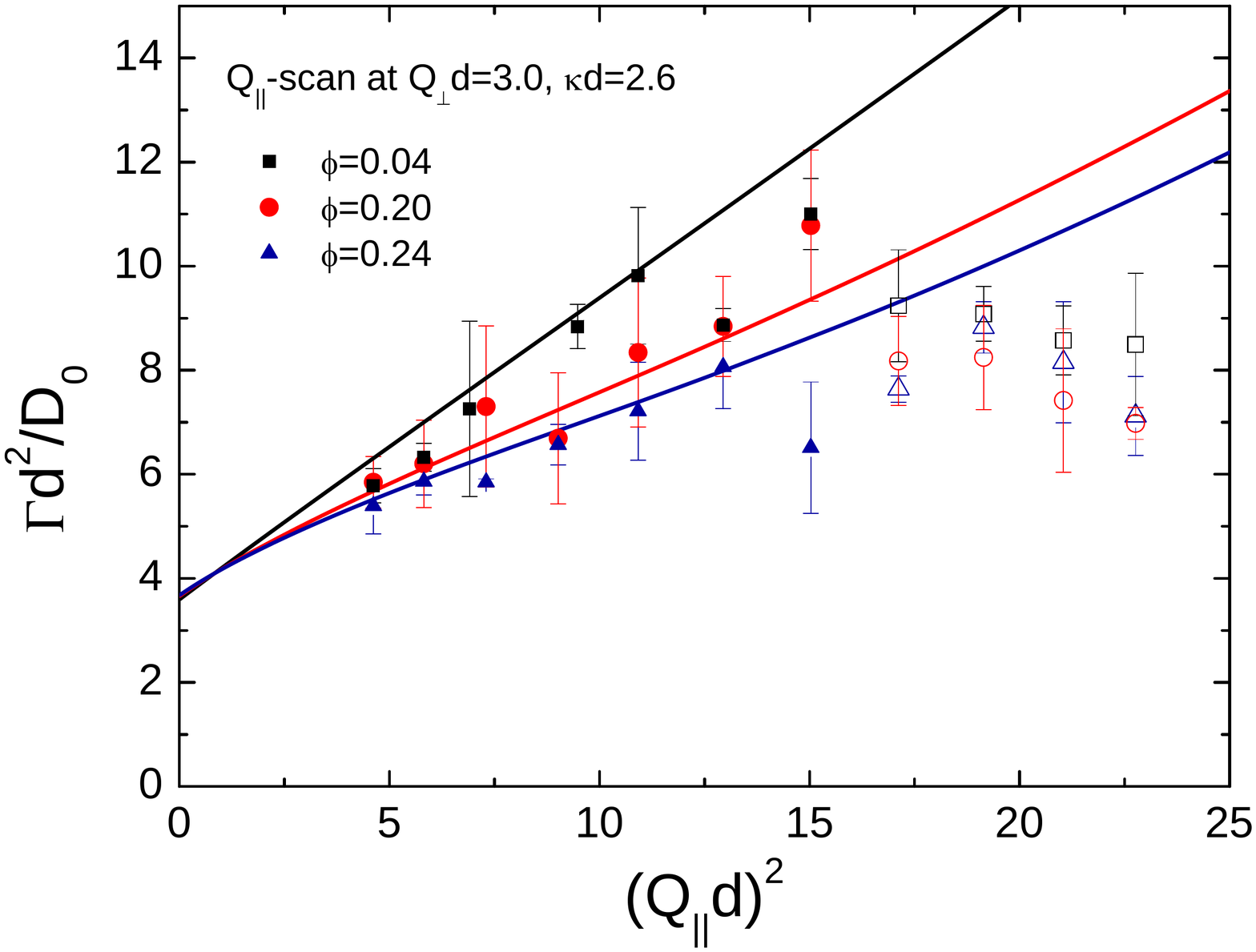}
   \caption{Relaxation rates versus parallel component of the scattering vector. Symbols represent experimental data obtained from ASM470 ($R_H=98$ nm) suspensions at different volume fractions at $Q_\perp d=3.0$ and $\kappa d=2.6$, lines are prediction by the virial approximation and open symbols refer to data points which are obscured by the back-reflection effect discussed in section \ref{sect:backrefl}}\label{fig:ex_virial_different_phi}
  \end{center}
\end{figure}

To illustrate the influence of the particle volume fraction, we display experimental data  of $\Gamma$ versus the scattering vector from $Q_\parallel$-scans with ASM470 suspensions at different volume fractions in Fig.  \ref{fig:ex_virial_different_phi}. It is obvious that at high $Q_\parallel$ the experimental data deviate from the virial approximation displayed as full lines in Fig. \ref{fig:ex_virial_different_phi} for all concentrations, which is fully explained by the effect of back-reflections, discussed in section \ref{sect:backrefl}. Apart from this high-$Q_\parallel$ deviation, the virial approximation predicts the experimentally observed at data correct, even at a sphere volume fraction of almost 25 percent.

\begin{figure}[ht]
  \begin{center}
 \includegraphics[width=0.8\columnwidth]{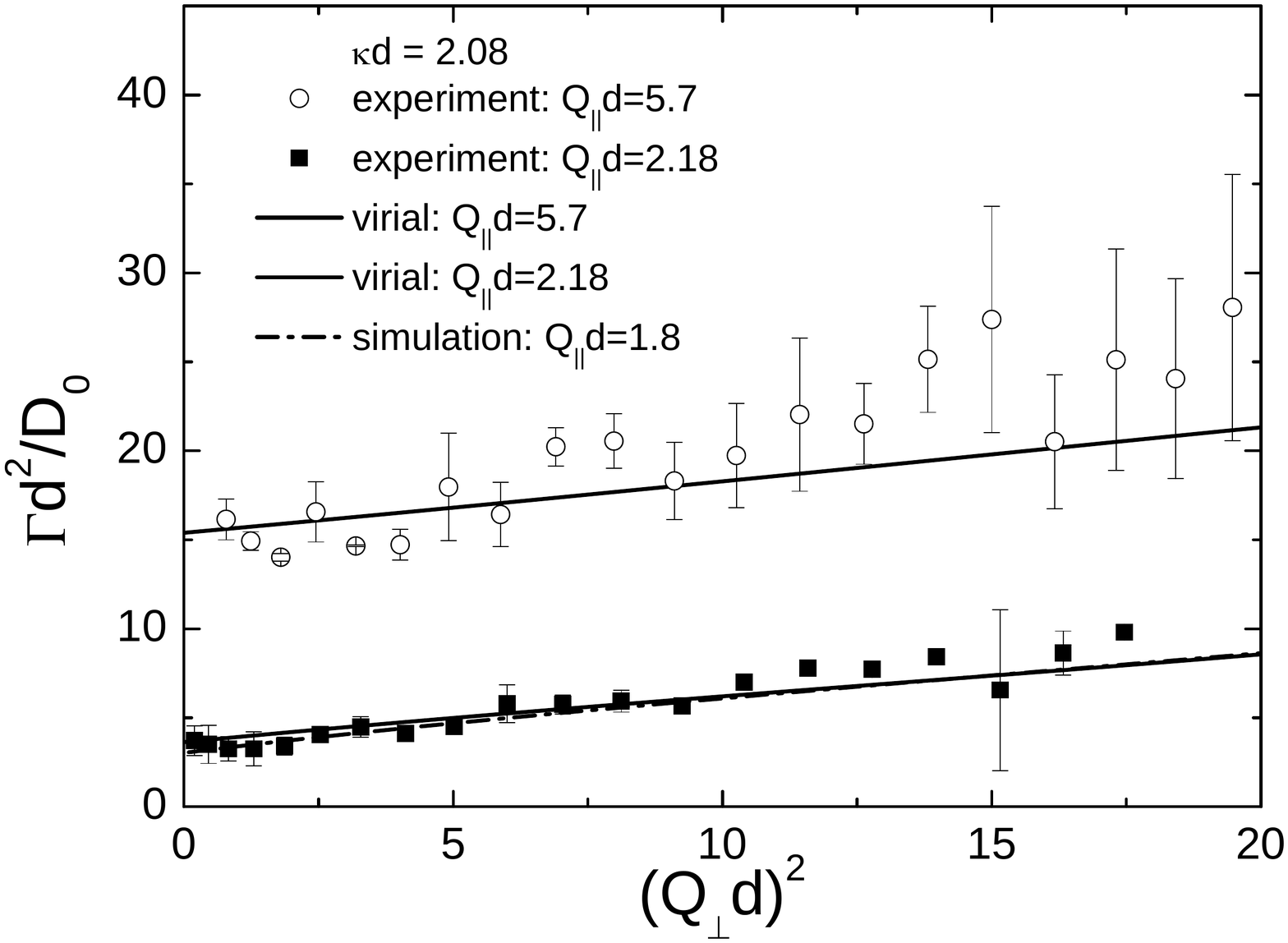}
   \caption{Relaxation rates versus normal component of the scattering vector. Symbols represent experimental data obtained from an ASM540 ($R_H=144$ nm) suspension with $\phi=0.175$ at $\kappa d=2.08$ at different values of the parallel scattering vector component, i.e. $Q_\parallel d=5.7$ (open circles) and $Q_\parallel d=2.18$ (full squares). Full lines are predictions by the virial approximation for the same experimental parameters and the dashed dotted line refers to simulation results obtained for $Q_\parallel d=2.18$ and $\kappa d= 1.8$.}\label{fig:ex_virial_different_qpd}
  \end{center}
\end{figure}

The same degree of agreement between virial approximation and experimental data is observed in $Q_\perp$-scans, which are shown in Fig. \ref{fig:ex_virial_different_qpd}. Here we display the experimental data obtained from the ASM540 suspension with $\phi=0.175$ performed at the same penetration depth $\kappa d=2.08$ but with extremely different values of parallel scattering vector component, i. e. $Q_\parallel=1.83$ and $Q_\parallel=5.7$. Together with these data we also present the results of BD-simulations which were obtained for a set of similar parameters, i. e. $Q_\parallel d=1.83$, $\kappa d =2.08$ and $\phi=0.15$. At low $Q_\parallel$, the results from all three methods agree very well, and at large $Q_\parallel$, where no simulation data are available, the agreement between experiment and virial approximation is also within the experimental error.

\begin{figure}[ht]
  \begin{center}
   \includegraphics[width=0.8\columnwidth]{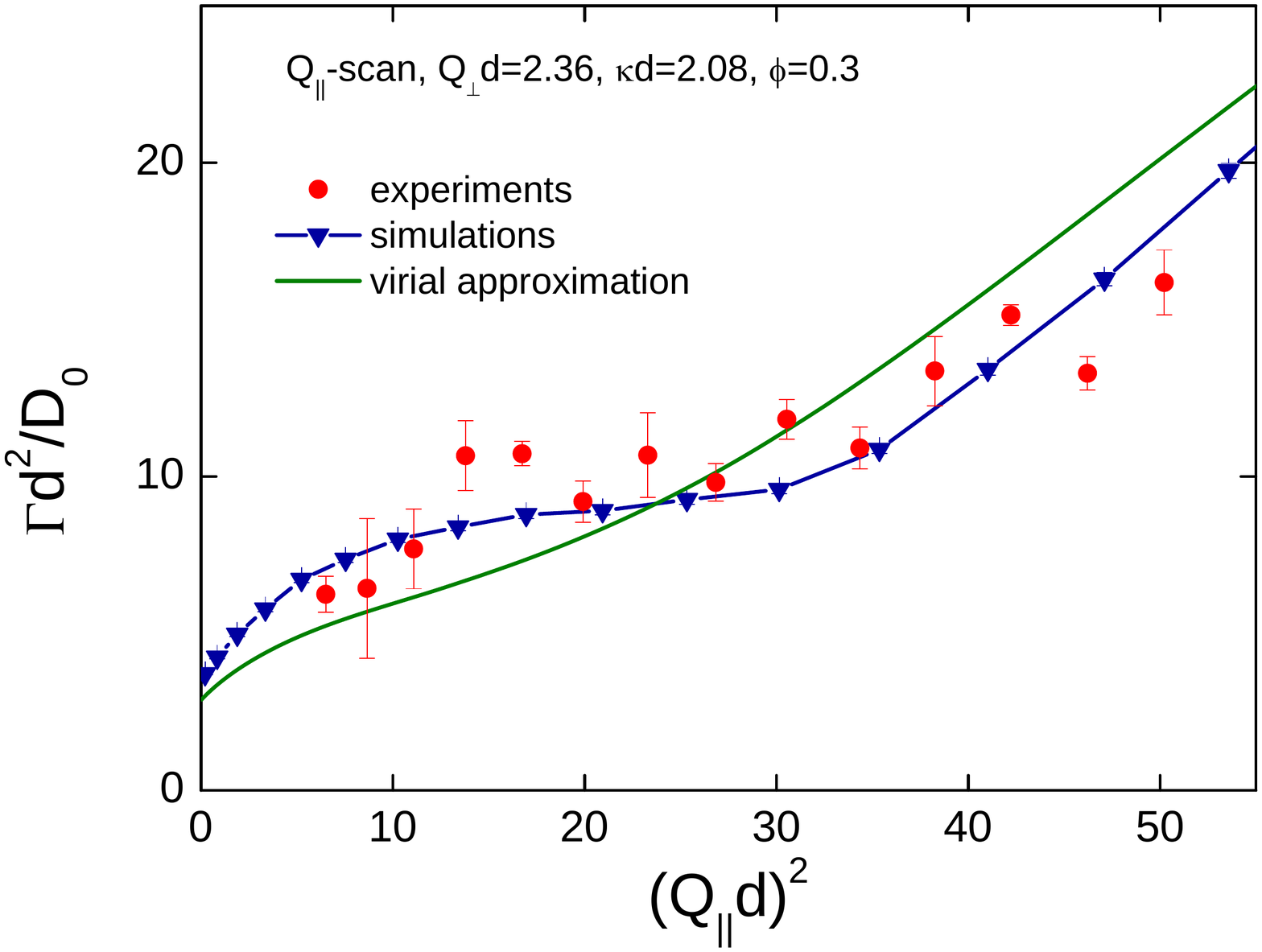}
   \caption{Relaxation rates versus parallel component of the scattering vector. Symbols represent experimental data obtained at $Q_\perp d=2.36$ and $\kappa d=2.08$ from an ASM540 ($R_H=144$ nm) suspension with $\phi=0.30$. The full lines are prediction by the virial approximation and the line with triangles refers to simulation results. Experimental data points, which are obscured by the back-reflection effect discussed in section \ref{sect:backrefl}, are omitted in this graph.}\label{fig:qpscan_phi30}
  \end{center}
\end{figure}

Only at the highest volume fraction ($\phi=0.3$) for which experimental data and simulations are available there is a significantly better agreement between simulation data and experiments than between virial approximation and measured data. This is shown in Fig.~\ref{fig:qpscan_phi30} where we display data from a $Q_\parallel$ scan, obtained from an ASM540 suspension with $\phi=0.3$ at $Q_\perp d=2.36$ and $\kappa d = 2.08$ together with  the corresponding predictions. At this high volume fraction the deviation between virial approximation and simulations is comparable or even larger than experimental error bars.

\begin{figure}[ht]
  \begin{center}
   \includegraphics[width=0.8\columnwidth]{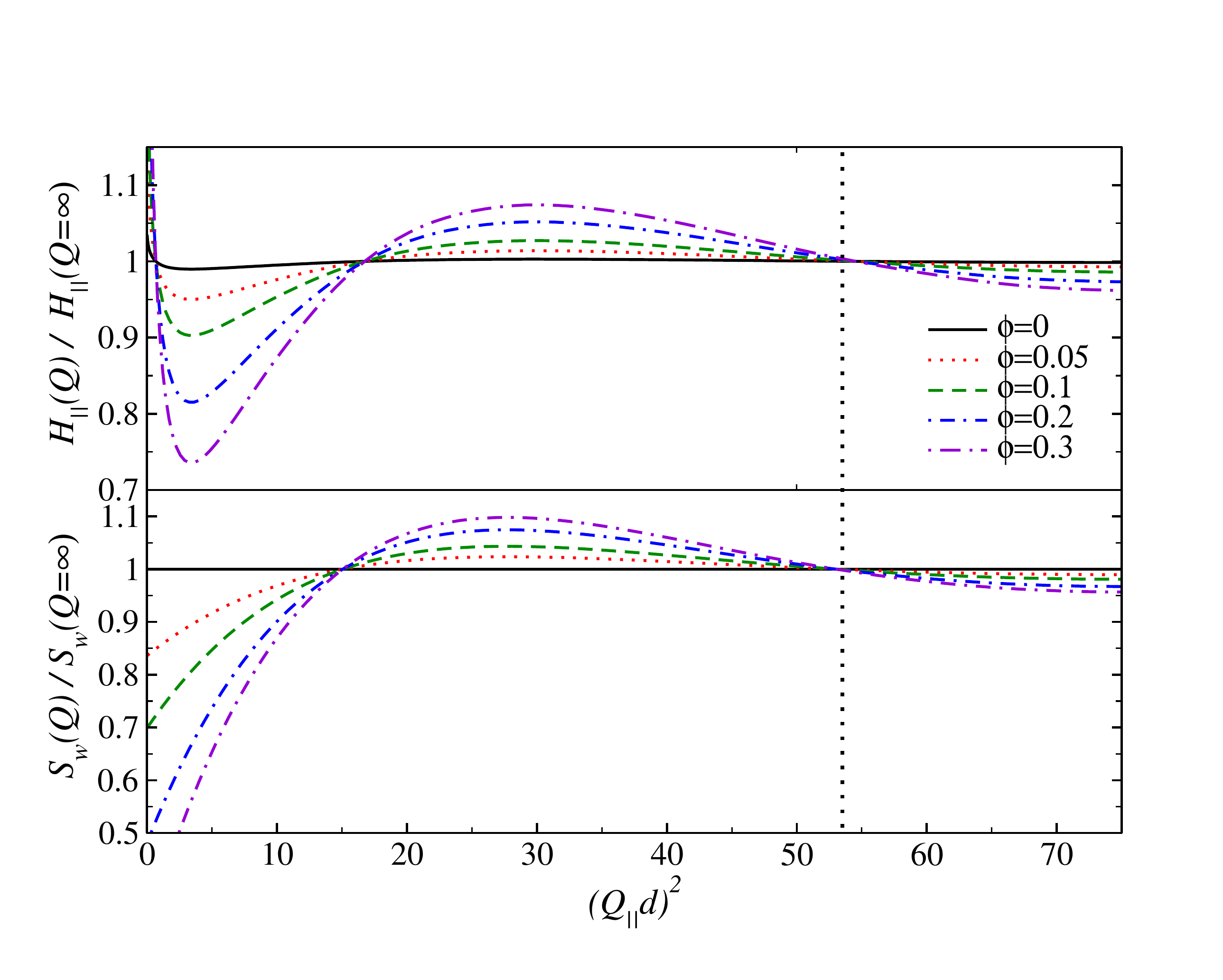}
   \caption{The wall structure factor $S_w(Q)$ and the hydrodynamic function $H_\parallel(Q)$ in a $Q_\parallel$-scan at fixed $Q_\perp d = 2.36$ and $\kappa d= 2.08$ for a selection of volume fractions. Both functions are normalised by their self-values at $Q\to\infty$. At $(Q_\parallel d)^2 \approx 53$ we find an isosbestic point for both functions (marked by the dotted vertical line), suggesting that the self-diffusion coefficients may be determined from the data collected in the vicinity of this point. The statement also holds sway for other components of $H$.}\label{isobestic}
  \end{center}
\end{figure}

It is interesting to investigate the limit of large scattering vectors, where self-diffusion is probed. As discussed in section \ref{sect:backrefl}, the relaxation rates determined at the largest scattering vectors (and thus the largest angle $\theta$) are not reliable. However, as first suggested by Pusey\cite{Pusey1978}, self-diffusion in bulk can be probed approximately at a wave vector $Q^*$ such that $S(Q^*)\approx S(Q\to\infty)$. This observation has been supported theoretically by Abade \textit{et al}\cite{Abade2011}. It is expected that at this point the distinct structure factor vanishes, and likewise does the distinct hydrodynamic function, so that only the self-parts contribute to the dynamic properties at this point. In the bulk case, this statement was later corroborated by extensive numerical simulations\cite{Segre1995,Banchio2008}. Segr\'e \textit{et al}\cite{Segre1995} stated that in a bulk suspension of hard spheres, this point is found for $Q^* a \sim 4.0$, where $S(Q^* a)=1$, to the right of the main peak of $S(\BQ)$. As shown by Banchio \textit{et al} \cite{Banchio2008b}, bulk structure factors of hard sphere suspensions with different volume fractions show an isosbestic point at $S(Qa=4.02)=1$ and at the same value of $Qa$ the corresponding hydrodynamic functions attain their high-Q limit. Michailidou {\it et al.}\cite{Michailidou2009} used the EWDLS experimental data at $Qa=4.58$ arguing that this should not be too far from $Q^*a$, thus providing a good estimate of the near-wall self-diffusion coefficient.Here, we propose a more thorough way to determine the particles' near-wall self-diffusion properties which follows the same line of arguments as discussed for bulk systems above. We note here that in EWDLS both the structure factor and the hydrodynamic function become penetration-depth dependent\cite{Cichocki2010}.
However, upon re-scaling by their asymptotic values, both $S_w(\BQ)$ and the components of $\BH_w(\BQ)$ exhibit an isosbestic point at which they attain their asymptotic values. We compute them using the virial expansion, and plot the results in Fig. \ref{isobestic}. Like for bulk experiments, first cumulants obtained at the $Q_{\parallel ,\perp}a$ values of the isosbestic point provide a good approximation for the near wall self-diffusion coefficients.

However, as the isosbestic point is found approximately at $Q^*_\parallel d=7.3$, we could determine experimental data of the first cumulant at this scattering vector only from the ASM540 suspensions. For the smaller ASM470 particles the data at $Q^*_\parallel d$ are in the range in which it is distorted by the back-reflection effect, and thus it may not be used to experimentally determine the self-diffusion coefficient parallel to the wall $\langle D_\parallel^s\rangle_\kappa$. Further, the experimentally accessible range of $Q_\perp d$ is in all cases much smaller than $Q^*_\perp d$ such that we can not get reliable experimental information on the self-diffusion properties normal to the wall.

\begin{figure}[ht]
  \begin{center}
    \includegraphics[width=0.8\columnwidth]{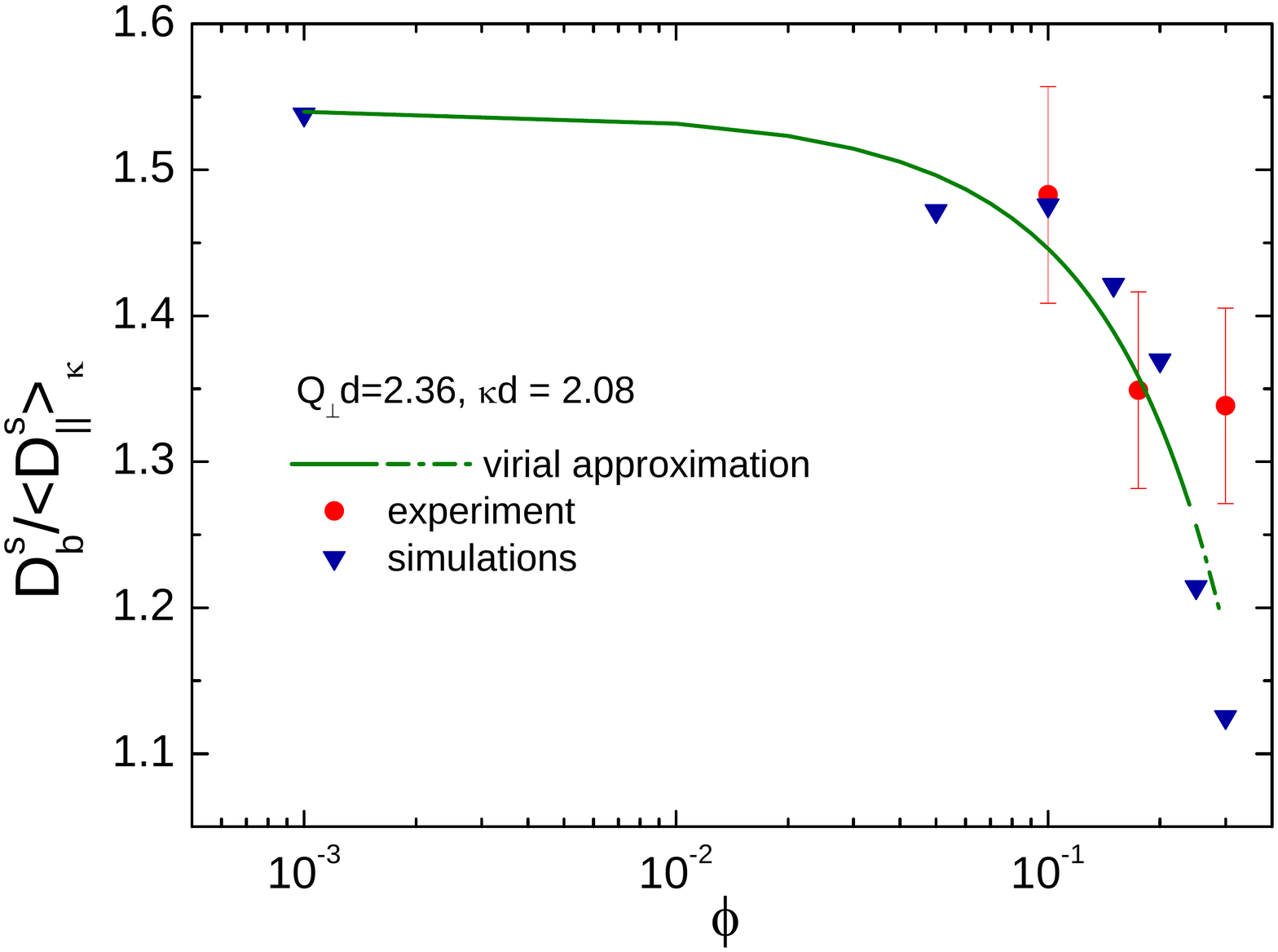}
   \caption{Comparison of experimental data (full circles) obtained from ASM470 ($R_H=98$ nm, $\phi=0.1$) and an ASM540 ($R_H=144$ nm), virial calculations (full line, the dashed dotted line represents virial calculations in a range of volume fractions where the approximation is not considered valid) and simulation results (triangles) for the self-diffusion coefficient parallel to the wall. Experimental parameters are at $Q_\perp d=2.36$ and $\kappa d=2.08$ for all cases.}\label{fig:d3_over_d2}
  \end{center}
\end{figure}

\begin{figure}[ht]
  \begin{center}
    \includegraphics[width=0.8\columnwidth]{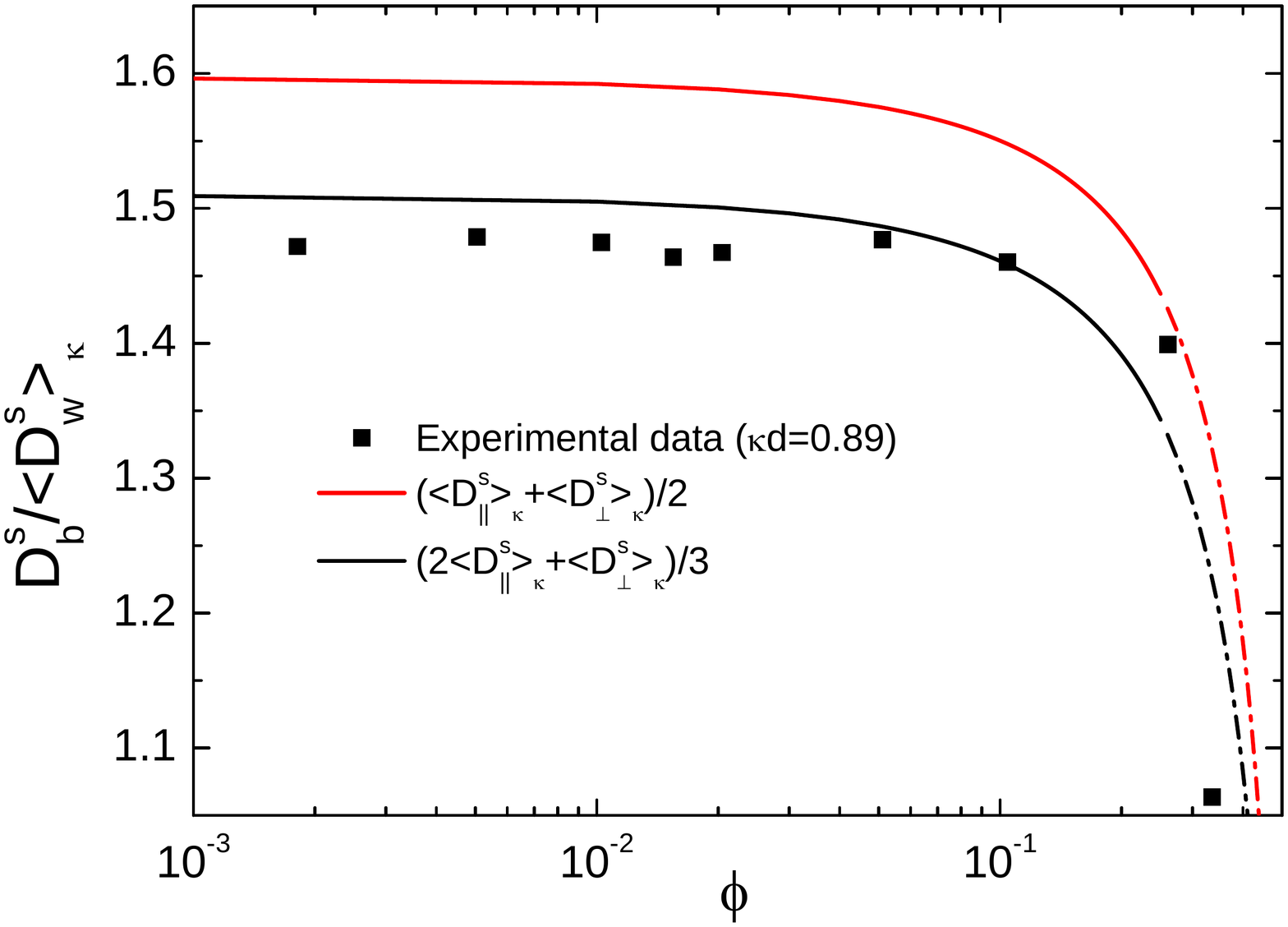}
   \caption{Comparison of virial predictions for the self-diffusion coefficient with experimental data by Michailidou et al~\cite{Michailidou2009} measured at $\kappa d=0.89$. The predicted data for normal and parallel contribution were averaged as indicated in the legend.}\label{fig:michailidou_data}
  \end{center}
\end{figure}

In Fig. \ref{fig:d3_over_d2}, we present the normalized ratios of $\langle D_{\parallel}^s\rangle$  over the bulk self-diffusion constant. The latter was calculated according to the semi-empirical formula\cite{Abade2011}
\begin{equation}
 \frac{D_b^s(\phi)}{D_0} = 1 -1.8315\phi (1+0.12\phi - 0.65\phi^2),
\end{equation}
which includes two virial coefficients due to Batchelor\cite{Batchelor1974} and Cichocki {\it et al.}\cite{Cichocki1999}, and is expected to be accurate up to $\phi\sim 0.45$. Its validity has been extended by Riest {\it et al.}\cite{Riest2015} up to $\phi=0.5$ by modifying the coefficient of the last term to $-0.70$.
We compare experimental data to predictions by virial approximation and simulations. The theoretical values for $\langle D_\parallel^s\rangle_\kappa$ were determined by linearly extrapolating the high-$Q$ range of the $\Gamma$ vs $Q_\parallel^2$ dependence, making use of Eq.~(\ref{Gamma_self}). Our experimental data confirm the trend predicted by both methods and show that the near-wall dynamics approach the bulk behaviour at high particle volume fractions. With this observation we qualitatively confirm the earlier results by Michailidou {\it et al.}\cite{Michailidou2009,Michailidou2013}.
Since the virial approach allows quick calculation of $\Gamma$ vs $Q_\parallel^2$ data, we can easily predict near wall self-diffusion coefficients for a variety of parameters, by using the slope in the high $Q$-range. We use this possibility to quantitatively compare self-diffusion properties predicted by the virial approximation to the data by Michailidou. For this purpose we calculate $\langle D_\parallel^s\rangle_\kappa$ and $\langle D_\perp^s\rangle_\kappa$ for a series of volume fractions and average them as $\langle D_w^s\rangle_\kappa=(\langle D_\parallel^s\rangle_\kappa + \langle D_\perp^s\rangle_\kappa)/2$ according to their experimental procedure. Their choice of $Qa=4.58$ is determined by the fact that they measured with a geometry which corresponds to $\theta=0^\circ$ and $\alpha_r=90^\circ$, thus at a scattering vector which makes an angle of $45^\circ$ with the interface. In this configuration the parallel contribution and the normal contribution to self-diffusivity are weighted equally in the experiment. The comparison in Fig.~\ref{fig:michailidou_data} shows that the prediction calculated by a $1:1$ weighing of the normal and the parallel component are deviating systematically from the experimental data in the range of volume fractions, where the virial approach should hold. Only at very high volume fractions, where the virial approximation is certainly not valid the experimental data appear to agree with it.
This is probably due to the effect that first cumulants obtained at $Qa=4.58$ are not a good approximation for the self-diffusion properties. Actually simulations of bulk properties\cite{Banchio2008b,Banchio2008b} show that even at moderate volume fractions, both the structure factor and the hydrodynamic function are significantly different from their value at $Q^*a=4.02$. For the sake of completeness we also show predictions for the self-diffusion constants in Fig.~\ref{fig:michailidou_data}, which are averaged according to $\langle D_w^s\rangle_\kappa=(2\langle D_\parallel^s\rangle_\kappa + \langle D_\perp^s\rangle_\kappa)/3$. These agree reasonably well with the earlier experimental data, which is probably a coincidence.

\begin{figure}[ht]
  \begin{center}
   \includegraphics[width=0.8\columnwidth]{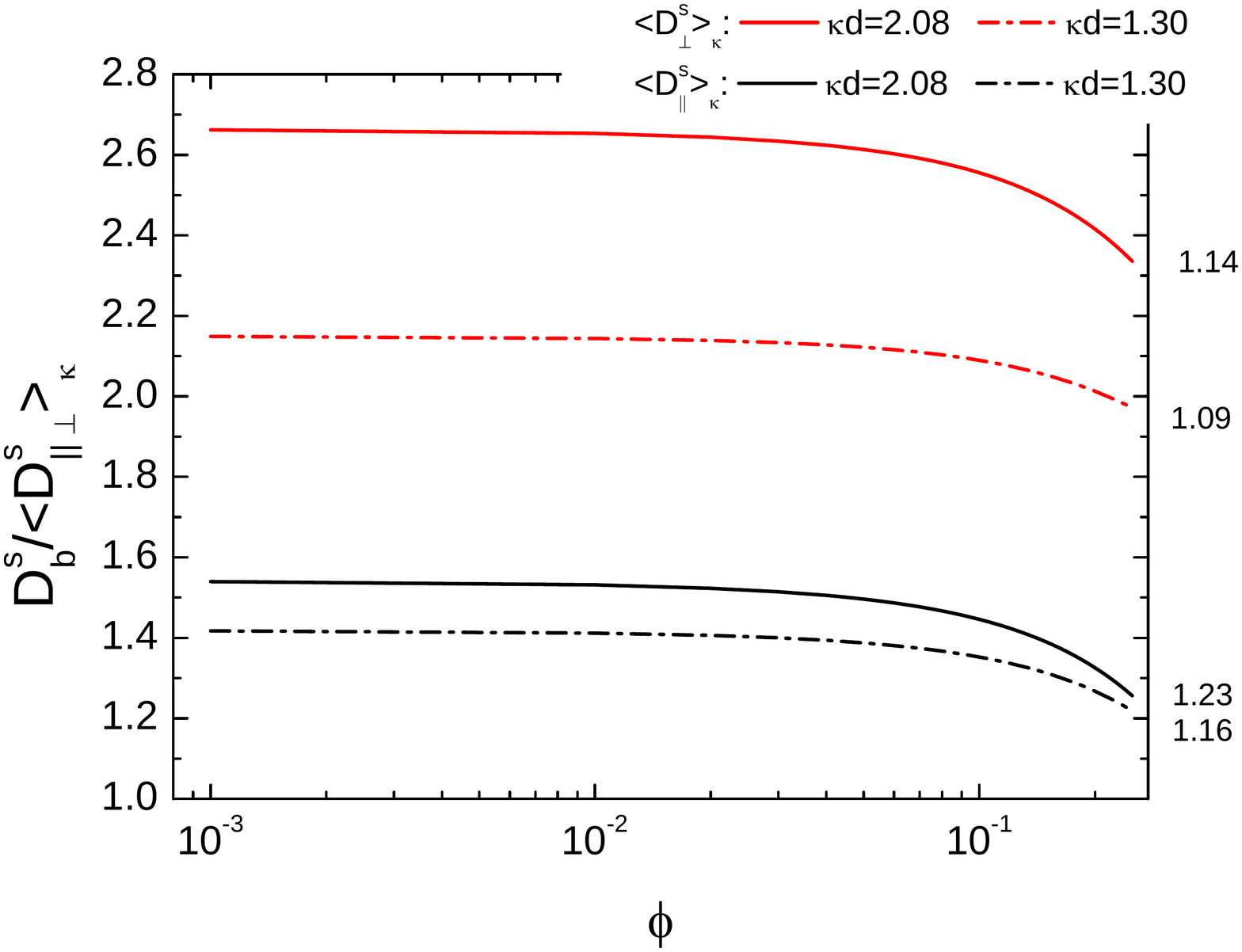}
   \caption{Calculated data for the self-diffusion coefficients parallel and normal to the wall for different penetration depths: $\kappa d=2.08$ full lines and $\kappa d=1.3$ dashed dotted lines. The numbers on the far right represent the ratio of the values at $\phi=10^{-3}$ over $\phi=0.25$, which are an indication that the diminishing of hydrodynamic interaction is more pronounced for particle motion parallel to the wall than normal to the wall.\label{fig:d3_over_d2_virial}}
  \end{center}
\end{figure}

Nevertheless, we confirm the earlier conjecture that particle-particle hydrodynamic interactions in the presence of a wall are diminished at high volume fractions as compared to bulk dynamics. However, here we can show that effect influences the diffusion parallel to the wall and normal to the wall differently. As discussed in Ref. \cite{Cichocki2010}, the anisotropic self-diffusion coefficients have the following virial expansion
\begin{equation}\label{self}
\frac{\avg{D^s_{\parallel,\perp}}_\kappa}{D_0} = G^{(1)}_{\parallel,\perp}(\kappa d) + \phi G^{(2)}_{\parallel,\perp}(\kappa d) + \mathcal{O}(\phi^2).
\end{equation}
\begin{table}[h]
\small
  \caption{\ The coefficients of the virial expansion of anisotropic self-diffusivity, defined in Eq. (\ref{self}). The decay of the $\perp$ elements is faster with increasing penetration depth, indicating that both single- and two-particle mobilities are hidered more for motion in the direction normal to the interface.}
  \label{tbl1}
  \begin{tabular*}{0.5\textwidth}{@{\extracolsep{\fill}}lllll}
    \hline
    $\kappa d$ & $G^{(1)}_{\perp}(\kappa d)$ & $G^{(2)}_{\perp}(\kappa d)$ & $G^{(1)}_{\parallel}(\kappa d)$ & $G^{(2)}_{\parallel}(\kappa d)$ \\
    \hline
 0	   &	  1.0    &  -1.832 & 1.0   & -1.832 \\
 0.2   &  0.781  &  -1.371 & 0.884 & -1.535 \\
 0.5   &  0.644  &  -1.117 & 0.810 & -1.357 \\
 1.0   &  0.516  &  -0.871 & 0.736 & -1.160 \\
 2.0   &  0.383  &  -0.588 & 0.654 & -0.903 \\
 5.0   &  0.227  &  -0.250 & 0.547 & -0.550 \\
    \hline
  \end{tabular*}
\end{table}
The coefficients of this expansion have been presented graphically in Fig. 3 of Ref.\cite{Cichocki2010}. Here we have tabulated them for a selection of penetration depths in Table \ref{tbl1}.  The coefficients have a clear interpretation: $G^{(1)}_{\parallel,\perp}$ refers to single-particle dynamics at infinite dilution, while $G^{(2)}_{\parallel,\perp}$ bears information on the effect of the wall on two-particle interactions. All coefficients decrease with increasing $\kappa d$, but the effect is stronger for the motion perpendicular to the wall. The behaviour of $G^{(1)}_{\parallel ,\perp}$ follows from the single-particle physical picture\cite{Lisicki2012}, in which motion normal to the interface is suppressed more than in the parallel direction. This is due to the fact that perpendicular motion generates 'squeezing' flows which lead to stronger hydrodynamic resistance as compared to 'shearing' flows induced by parallel motion\cite{Kim}. The particle-particle HI are affected in the same way, which explains the faster decay of $G^{(2)}_\perp$ as compared to $G^{(2)}_\parallel$. Thus, the coefficients corresponding to the normal motion are affected more strongly. However, the near-wall self-diffusivity is frequently written in the form
\begin{equation}
\frac{\avg{D^s_{\parallel,\perp}}_\kappa}{\avg{D_{\parallel,\perp}}_\kappa } = 1 - \alpha_{\parallel,\perp}(\kappa d) \phi + \ldots,
\end{equation}
with $\avg{D_{\parallel,\perp}}_\kappa =D_0 G^{(1)}_{\parallel,\perp}(\kappa d)$. The coefficient
\begin{equation}
\alpha_{\parallel,\perp}(\kappa d)= \frac{ G^{(2)}_{\parallel,\perp}(\kappa d)}{ G^{(1)}_{\parallel,\perp}(\kappa d)},
\end{equation}
becomes a result of an interplay between the single- and two-particle effects.  In Fig.~\ref{fig:d3_over_d2_virial} we show normalized ratios of $\langle D_\parallel^s\rangle_\kappa$ and $\langle D_\perp^s\rangle_\kappa$ over the bulk self-diffusion as a function of volume fractions for two different penetration depths of the evanescent wave. The curves are calculated using the virial approach up to a volume fraction of 25\%. First we observe that the self-diffusion coefficient (averaged over the illumination profile) normal to the wall is smaller than that parallel to the wall and that both components increase with penetration depth $\kappa^{-1}$, similarly to the components of the near-wall diffusion coefficients at infinite dilution. The variation of these ratios over the range of volume fractions covered is indicated by the numbers on the far right of Fig.~\ref{fig:d3_over_d2_virial}, which are the ratios of the values obtained at $\phi=10^{-3}$ and $\phi=0.25$. It is important to note, that although $\avg{D^s_\parallel}_\kappa/D^s_b$ varies stronger with increasing $\phi$ as compared to $\avg{D^s_\perp}_\kappa/D^s_b$, this does not imply that the wall diminishes the particle-particle HI more in the parallel direction, as we discussed above.

\section{Conclusions}\label{sect:conclusions}

In this paper we describe our EWDLS investigations of the near wall dynamics of colloidal hard spheres in suspensions with volume fractions up to $\phi=0.3$ We thoroughly compare experimental data for the dependence of the first cumulant on the scattering vector components parallel and normal to the interface to corresponding predictions based on a second order virial approximation and to simulation results, where the full hydrodynamic interaction is taken into account. Up to volume fractions of about fifteen to twenty percent we find perfect agreement between the three methods. Above this range, the predictions by the virial approach deviate discernibly from the simulation data~\cite{Cichocki2010}, however this deviation is still in the range of experimental error bars. Therefore we conclude that the virial approach provides a good approximation for the prediction and analysis of experimental data up to a volume fraction of about 25\%, which is much less time consuming and elaborate than full scale simulations. Only at $\phi\geq 0.3$ the virial approximation is clearly not anymore able to capture the details of the dependence of the first cumulant on the scattering vector.
Further we introduce a new method to assess the particles' near wall self-diffusivity from experimental data. This method follows the same line of argument, which is used to assess bulk self-diffusivity in cases where the limit of sufficiently high scattering vector cannot be reached experimentally. We identify an isosbestic point of the near-wall structure factors right to the first maximum, where near wall structure factor and hydrodynamic function attain their asymptotic values. Diffusion data measured at the scattering vector of the isosbestic point are a good approximation for the self-properties. Comparison of experimental data with predictions, based on the virial approach and on simulations, show that this method yields better estimates of the self-diffusivity as methods used earlier.
Finally we confirm earlier data which show that the diminishment of particle-particle hydrodynamic interactions
due to the presence of the wall is less pronounced at high volume fraction compared to bulk dynamics. Beyond that, we show (see Table \ref{tbl1}) that the
observed effect is weaker for the mobility parallel to the wall as compared to motion in the normal direction.
In conclusion, with the virial approximation, we have a method at hand, which qualitatively supports earlier data, but provides significant further insight into the near wall dynamics of colloidal hard spheres. This is especially important since this approach can be easily adopted to systems with long ranging static interaction, providing a quick and non-costly method for the prediction and analysis of EWDLS results obtained from e. g. charged colloids.

\section*{Acknowledgements}
YL thanks for support from Marie Sklodowska Curie initial Training network SOMATAI under EU Grant Agreement No. 316866. J.B. was supported by NSF Grant No. CBET-1059745.  He would also like to acknowledge the hospitality and financial support from IPPT PAN during his summer visits. ML wishes to acknowledge support from the National Center of Science grant no. 2012/07/N/ST3/03120. EW was supported by the the National Center of Science grant no. 2012/05/B/ST8/03010. The work was supported by the Foundation for Polish Science International PhD Projects Programme co-financed by the EU European Regional Development Fund. Part of the experimental data presented was obtained with financial support from the European Commission under the Seventh Framework Program by means of the grant agreement for the Integrated Infrastructure Initiative No. 262348 European Soft Matter Infrastructure (ESMI) which is gratefully acknowledged.

%%%REFERENCES%%%
%\bibliography{rsc} %You need to replace "rsc" on this line with the name of your .bib file
%\bibliographystyle{rsc} %the RSC's .bst file

\end{document}